\documentclass[11pt]{article}
\usepackage{amsmath}
\usepackage{amssymb}
\newtheorem{lemma}{Lemma}

\newtheorem{thm}{Theorem}

\numberwithin{equation}{section}
\newtheorem{note}{Note}

\begin{document}

\title{The Group Classification of One Class \\
of Nonlinear Wave Equations.}

\author{ V.
Lahno\\ \small Pedagogical University, 2 Ostrohradskyj Street,
314000 Poltava, Ukraine\thanks{e-mail:
laggo@poltava.bank.gov.ua}
\and O.~Magda \\ \small Institute of
Mathematics,  3 Tereshchenkivska Street, 252004 Kyiv,
Ukraine\thanks{e-mail: magda@imath.kiev.ua}}

\date{}

\maketitle

\begin{abstract}
The problem of group classification of one class of quasilinear
equations of hyperbolic type with two independent variables has
been solved completely.

\end{abstract}

\section{Introduction}

The problem of group classification of differential equations is
one of the central problems of  modern symmetry analysis of
differential equations \cite{magda1}. One of the important classes
are hyperbolic equations. The problem of group classification of
such equations has been discussed by many authors (see
e.g.[1--8],[10--11]).
\begin{eqnarray*}
\cite{magda32,magda43,magda9} && u_{tt} = u_{xx} +F(u) ;\\
\cite{magda10,magda35,magda11} && u_{tt} = [f(u)u_{x}]_x; \\
\cite{magda11,magda12} && u_{tt} = f(u_x) u_{xx};\\
 \cite{magda31}  && u_{tt} = F(u_x) u_{xx} +H(u_x);\\
\cite{magda11} && u_{tt} = F(u_{xx}) ;\\
 \cite{m79} && u_{tt} = u^m_x u_{xx}+f(u) ;\\
\cite{m80} && u_{tt}+f(u) u_t = (g(u) u_x)_x +h(u) u_x .
\end{eqnarray*}

 In this article we consider  a classes of hyperbolic
equations in $1+1$ time-space:

\begin{eqnarray}
u_{tx}& =& g(t,x) u_x +f(t,x,u), \ g_x \not =0, \ f_{uu} \not =0;
\label{eq1}\\
 u_{tx} &=& f(t,x,u), \ \ f_{uu} \not =0. \label {L3}
 \end{eqnarray}
 where \ $u=u(t,x)$ and $ g,f$ is an arbitrary nonlinear differentiable function,
is an arbitrary nonlinear smooth function, which dependent
variables \ $u$ \ or \ $u_x$. We use following notation \
$u_x=\displaystyle{\frac{\partial u}{\partial x}}, \
u_{xx}=\displaystyle{\frac{\partial ^2u}{\partial x^2}}, \
u_t=\displaystyle{\frac{\partial u}{\partial t}}, \ .$ The
approach used in the present article is that presented in
\cite{zhd99}, being a synthesis of the standard Lie algorithm for
finding symmetries and the use of canonical forms of partial
differential operators obtained with the equivalence group of the
equation at hand. The conditions for a linear partial differential
operator to be a symmetry operator of equation (\ref{eq1}) turn
out to be too general to produce a manageable system of defining
equations. A solution to this apparent impasse is to invoke the
equivalence group of equations (\ref{eq1}), (\ref{L3}) and use it
to find canonical forms for the symmetry operators. In
\cite{zhd99} this method was applied to nonlinear heat equations.

The method is as follows. First, establish the defining system for
a partial differential operator to be a symmetry operators of
equations (\ref{eq1}),(\ref{L3}). Then calculate the equivalence
group of the equation. Here, the equivalence group is following
group of invertible transformations

\begin{equation}\label{equitran}
\bar{t}=T(t,x,u),\quad \bar{x}=X(t,x,u),\quad \bar{u}=U(t,x,u)
\end{equation}
which transforms equations (\ref{eq1}),(\ref{L3}) to  equations of
the same forms

\begin{eqnarray}
\bar{u}_{\bar{t}\, \bar{x}} &=& \bar{u}_{\bar{x}\, \bar{x}}
+\bar{g}(\bar{t},\bar{x}) \bar{u}_{\bar{x}} +
\bar{f}(\bar{t},\bar{x},\bar{u}); \label{eqbis}\\
 \bar{u}_{\bar{t}\, \bar{x}} &=&
\bar{f}(\bar{t},\bar{x},\bar{u}).\label{eqbis1}
\end{eqnarray}
The next step is to calculate the various canonical forms for a
linear partial differential operator (LPDO) with respect to this
equivalence group. This is the same as linearizing the operator
(more precisely, finding a simplest form of an arbitrary LPDO
which is equivalent under the equivalence transformations
(\ref{equitran}) to the LPDO at hand).

However, these canonical forms for LPDOs and symmetry operators on
their own provide no solution to the problem we wish to solve: we
do not, unlike for the cases solvable by Lie's algorithm,
automatically obtain the Lie algebra of symmetry operators in this
way. Our procedure requires us to make assumptions about the
nature of the Lie symmetry algebra of equations (\ref{eq1}),
(\ref{L3}). We take an arbitrary Lie algebra and find, using our
equivalence group and the conditions for an LPDO to be a symmetry
operator of equations (\ref{eq1}), (\ref{L3}) canonical
representations of the Lie algebra as a symmetry algebra of
equations (\ref{eq1}),(\ref{L3}). Different representations will
give different forms for the functions $f(t,x,u),\; g(t,x,u).$ At
present we have no result which tells us the maximal dimension for
the Lie symmetry algebra, unlike the case of ordinary differential
equations. However, we find that we need only Lie algebras of
dimension no greater than three.

 All our arguments are local, and we do not treat global questions.
  Also, all  functions involved in our arguments are assumed to be
   continuously differentiable of the appropriate order.

  The first step in our programme is to find the conditions for a
vector field to be a point symmetry of equations (\ref{eq1}),
(\ref{L3}). To this end we consider a vector field of the form

\begin{equation}\label{eq2}
Q= \tau(t,x,u)\partial_t + \xi(t,x,u)\partial_x +
\eta(t,x,u)\partial_u
\end{equation}
where\ $\tau $,\ $\xi $,\ $\eta $ are arbitrary, real-valued
smooth functions defined in some subspace of the space\
$V=X\otimes R^1$\ of the independent variables\ $X=\langle t,x
\rangle $\ and the dependent variable\ $R^1=\langle u \rangle$. As
a result, we find that the operator (\ref{eq2}) generates a
one-parameter symmetry group of equation (\ref{eq1}) if
\begin{equation}
\begin{split}
&\varphi^{tx}-\varphi^{xx}-[\tau g_t-\xi g_x]u_x-\varphi^x g -\tau
f_t-\xi f_x-\eta f_u \Big|_{u_{tx}=gu_{x}+f}=0,
\end{split}
\end{equation}
where
\begin{eqnarray}
\varphi ^t  &=& D_t(\eta )-u_tD_t(\tau )-u_xD_t(\xi ), \nonumber
\\ \varphi ^x  &=& D_x(\eta )-u_tD_x(\tau )-u_xD_x(\xi ),
\nonumber \\ \varphi ^{xx}&=& D_x(\varphi ^x)-u_{tx}D_x(\tau
)-u_{xx}D_x(\xi ) \nonumber\\
\end{eqnarray}
and\ $D_t$,\ $D_x$ are operators of total differentiation in $t$
and $x$ respectively. \\
We then find the following results:
\begin{thm} The symmetry group of the equation  (\ref{eq1})
is generated by the infinitesimal operators of the form
\begin{equation} \label{n1}
Q = \tau(t) \partial_t+\xi(x) \partial_x+[h(t)u
+r(t,x)]\partial_u,
\end{equation}
where\  $ \tau,\  \xi, \  \eta$ \  are real-valued functions that
satisfy the system
\begin{eqnarray} \label{1.1}
&& r_{tx} +f[h-\tau_t-\xi_x] = g r_x +\tau f_t +\xi f_x +[hu+r]f_u , \\
&& h_t = \tau_t g+\tau g_t +\xi g_x. \nonumber
\end{eqnarray}
\end{thm}
\begin{thm} The symmetry group of the equation  (\ref{L3})
is generated by the infinitesimal operators of the form
\begin{equation} \label{n11}
Q = \tau (t ) \partial_t+\xi(x) \partial_x+(ku+r(t,x))\partial_u,
\end{equation}
where\  $ \tau,\  \xi, \  \eta$ \  are real-valued functions that
satisfy the equation
\begin{eqnarray} \label{m1}
r_{tx} +[k-\tau'-\xi']f=\tau f_t +\xi f_x +[ku+r]f_u. \nonumber
\end{eqnarray}
\end{thm}
\begin{lemma}
The maximal equivalence group\ $\cal E$\  of equation (\ref{eq1})
 reads as
\begin{eqnarray}\label{1.2}
(1) && \bar t = T(t) , \ \bar x = X(x), \ v = U(t)u +Y(t,x), \ \ t'X'U \not =0;  \\
(2) && \bar t = T(x) , \ \bar x = X(t), \ v = \Psi(x)\Phi(t,x)u +Y(t,x), \ \ t'X'\Psi \not =0, \nonumber\\
&& \Phi(t,x) = \exp[-\int g(t,x) dt], \ g_x \not=0. \nonumber
\end{eqnarray}
where\ $\dot T \not =0,$ \ $\displaystyle{\frac{D(X,U)}
{D(x,u)}}\not =0$.

\end{lemma}
\begin{lemma}
The maximal equivalence group\ $\cal E$\  of equation (\ref{L3})
 reads as
\begin{eqnarray}\label{E1}
(1) && \bar t = T(t), \ \ \bar x = X(x), \ \ v = mu +Y(t,x), \\
(2) && \bar t = T(x), \ \ \bar x = X(t), \ \ v=mu +Y(t,x), \ \ T'
X' m \not =0. \nonumber
\end{eqnarray}
where\ $\dot T \not =0,$ \ $\displaystyle{\frac{D(X,U)}
{D(x,u)}}\not =0$.

\end{lemma}

The proof is by direct calculation using the chain rule, and we
omit it here.

The first step in our method is to take a canonical form for a
vector field. This is essentially the same as linearising a vector
field, but we use only the transformations of the equivalence
group $\mathcal{E}$ of the equations (\ref{eq1}), . The reason for
this is that the linearising transformations must not take us out
of the classes of equations of the type given in (\ref{eq1}),
(\ref{L3}). We characterize the possible canonical forms for
vector fields in the following result:

\begin{thm}\label{1.3}
There are changes of variables (\ref{1.2}) that reduce an operator
(\ref{n1}) to one of the operators below:
\begin{eqnarray} \label{1.4}
Q&=& t \partial_t+x\partial_x; \ \ Q = \partial_t; \ \ Q = \partial_x+tu\partial_u; \nonumber \\
Q&=& \partial_x+\epsilon u \partial_u, \ \epsilon =0,1; \ Q = tu \partial_u, \\
Q&=& u \partial_u, \ \ Q = r(t,x) \partial_u, \ r \not =0.
\nonumber
\end{eqnarray}
\end{thm}
{\bf Proof.} The first group of transformation  (\ref{1.2})
reduces operator $Q$ (\ref{n1}) to the form
\begin{equation} \label{1.5}
\tilde Q = \tau T' \partial_{\bar t} +\xi X' \partial_{\bar x}
+[(\tau U' +Uh) u+\tau Y_t+\xi Y_x +Ur] \partial_v.
\end{equation}
If $\sigma \cdot \xi \not =0,$ then, the function $T, X, U$
(\ref{1.2}) is a solutions of the first-order PDE
$$ \tau T' = T, \ \ \xi X'= X, \ \ \tau U' +h U=0, $$
and the function $Y$ is a solution of the equation
$$ \tau Y_t +\xi Y_x +Ur =0,$$
we find that the operator  $\tilde Q$ (\ref{1.5}) takes the form
$$ \tilde Q = \bar t\partial_{\bar t} +\bar x \partial_{\bar x}.$$
If $\tau \not =0,$ a $\xi =0,$ choosing in (\ref{1.2}) $T, U, Y$ a
particular solution of PDE
$$ \tau T' =1, \ \tau U' +h U=0 \ (U \not =0), \ \tau Y_t +U r=0,$$
we transform (\ref{n1}) to become
$$\tilde Q = \partial_{\bar t}.$$
If $\tau =0, \ \xi \not =0,$ and
\begin{itemize}
\item $h' \not =0$ we get the operator $\tilde Q =\partial_{\bar
x} +\bar t v\partial_v.$
\item $h'=0,$
 we get the operator$\tilde Q =
\partial_{\bar x} +\epsilon v \partial_v,$ where $\epsilon =0$ or
$\epsilon =1.$
\end{itemize}
If $\tau = \xi =0,$ that we get the following case:
 $$ \tilde Q = \bar t v\partial_v, \ \ \tilde Q =v \partial_v, \ \ \tilde Q = r(\bar t, \bar x) \partial_v.$$
The theorem is proved.

\begin{thm}\label{1.32}
There are changes of
variables (\ref{1.23}) that reduce an operator (\ref{n11}) to one
of the operators below:
\begin{eqnarray*}
Q&=& \partial_t +\partial_x+\epsilon u \partial_u \ \ (\epsilon =0,1):\\
Q&=& \partial_t +\epsilon u \partial_u \ \ (\epsilon =0,1); \\
Q&=& u \partial_u, \ Q = g(t,x) \partial_u \ (g \not =0).
\nonumber
\end{eqnarray*}
\end{thm}

\begin{thm}\label{t2}
There are three equations of the form given in (\ref{eq1}) which
admit local one-parameter symmetry groups generated by the
canonical forms given in Theorem \ref{1.3}. They are described by
the following list, where $\langle Q\rangle$ denotes the algebra
generated by the operator $Q$ and we define the equation of the
form (\ref{eq1}) by the form of the functions $g(t,x)$ and
$f(t,x,u)$:
\begin{eqnarray*}
A^1_1 &=& \langle t \partial_t +x \partial_x \rangle : g = t^{-1} \tilde g(\omega), \ f = t^{-2} f(u, \omega), \ \ \omega = tx^{-1}, \\
&& \tilde g_\omega \not =0, \ \ f_{uu} \not =0; \\
A^2_1& =& \langle \partial_t \rangle: g = \tilde g(x), \ f = \tilde f(x,u), \ \tilde g' \not =0, \ \tilde f_{uu} \not =0; \\
A^3_1 &=& \langle \partial_x +tu \partial_u \rangle : g=x+\tilde
g(t), f = e^{tx} \tilde f(t,\omega), \ \omega = e^{-tx} u, \
\tilde f_{\omega \omega} \not =0.
\end{eqnarray*}
\end{thm}
 Having established the one-dimensional Lie point symmetry algebras of
 equation (\ref{eq1}) we deal with
the semi-simple Lie algebras. In fact, we show that no semi-simple
Lie algebra has a representation
 in terms of the given vector fields. We prove this  result for the two real simple Lie algebras
 $ so(3)$ and $ sl(2,R ).$

 \begin{thm} The real simple Lie algebras $ so(3)$ and
$ sl(2, R)$ do not have any realizations as symmetry algebras of
equation (\ref{eq1}).
\end{thm}

\smallskip\noindent{\bf Proof:} First, $ so(3).$ The commutation
relations are

\[
[e_1, e_2]= e_3,\;\; [e_2, e_3]= e_1,\;\; [e_3, e_1]= e_2.
\]
One of these operators may be taken to be in one of the canonical
forms given in Theorem \ref{1.3}. We do the calculation for the
first canonical form and take $e_1=t\partial_t+ x\partial_x.$ Then
we take $e_2$ and $e_3$ in general form:

\begin{eqnarray*}
e_2 &=& \lambda e_1 + \lambda_1\partial_t + \lambda_2\partial_x +(h(t)u+r(t,x))\partial_u \\
e_3 &=& \mu e_1 + \mu_1\partial_t + \mu_2\partial_x
+(g(t)u+s(t,x))\partial_u.
\end{eqnarray*}
We may set $\lambda = \mu =0$ since $e_1$ commutes with itself
(this is equivalent to replacing $e_2$ and $e_3$ by $e_2-\lambda
e_1$ and $e_3-\mu e_1)$. Thus we take

\begin{eqnarray*}
e_2 &=& \lambda_1\partial_t + \lambda_2\partial_x +(h(t)u+r(t,x))\partial_u \\
e_3 &=& \mu_1\partial_t + \mu_2\partial_x
+(g(t)u+s(t,x))\partial_u.
\end{eqnarray*}
Then $[e_1, e_2]=e_3$ and $[e_3, e_1]= e_2$ give us

\begin{equation*}
\begin{split}
\mu_1\partial_t + \mu_2\partial_x +(g(t)u+s(t,x))\partial_u&=
-\lambda_1\partial_t - \lambda_2\partial_x +\\
& + (th'(t)u+tr_t(t,x)+xr_x(t,x))\partial_u´\\
\end{split}
\end{equation*}

\begin{equation*}
\begin{split}
-\lambda_1\partial_t - \lambda_2\partial_x
-(g(t)u+s(t,x))\partial_u&=-
\mu_1\partial_t - \mu_2\partial_x +\\
& +(tg'(t)u+ts_t(t,x)+xs_x(t,x))\partial_u
\end{split}
\end{equation*}
From this we see that $\lambda_1=\lambda_2=\mu_1=\mu_2=0$ so that

\[
e_2=(h(t)u+r(t,x))\partial_u,\;\; e_3=(g(t)u+s(t,x))\partial_u
\]
which gives $[e_2, e_3]=0\neq e_1.$ All the other calculations
lead to the same result for both $ so(3)$ and $ sl(2, R).$ \\
The theorem is proved.\\
From this theorem we obtain  the following :
\begin{note}
In the class of operators (\ref{eq1}) there are no realizations of
any real semi-simple Lie algebras;
\end{note}
\begin{note}
 There are no  equations (\ref{eq1}) which has algebras of invariance, which are isomorphic
by real semi-simple algebras, or conclude those algebras as
subalgebras.
\end{note}

Every solvable Lie algebra ${\sf g}$  has, as is well-known, a
composition series

\[
{\sf g}={\sf g}_n\rhd {\sf g}_{n-1}\rhd \ldots \rhd {\sf
g}_{0}=\{0\}
\]
where each ${\sf g}_i$ is an ideal of codimension one in ${\sf
g}_{i+1}$ for $i=0,\ldots, n-1.$ This structure allows us to
perform an inductive construction of the realizations of solvable
Lie algebras as vector fields: having constructed  realizations
for solvable Lie algebras of dimension $n$, we may construct a
realization for any solvable Lie algebra  ${\sf g}_{n+1}$ of
dimension $n+1$ by adding the appropriate $(n+1)$-st element to a
solvable Lie algebra ${\sf g}_n$ of dimension $n$ in the
composition series of ${\sf g}_{n+1}.$

\medskip
In Theorem \ref{t2} we gave all realizations of one-dimensional
Lie algebras which give symmetries of equation (\ref{eq1}): there
are three inequivalent such realizations. We then note that there
are two inequivalent solvable Lie algebras of dimension two with
generators $e_1, e_2:$

\[
\begin{split}
 A_{2.1}&=\langle e_1, e_2\rangle,\quad [e_1, e_2]=0\\
A_{2.2}&=\langle e_1, e_2\rangle,\quad [e_1, e_2]=e_2,
\end{split}
\]

\begin{thm}\label{t3}
Any $A_{2.2}$-invariant equation of the type (\ref{eq1})
is equivalent
 to one of the following inequivalent equations (where we give the realization
 of $A_{2.2}$ and the forms of the functions $f$ and $g$).
\begin{eqnarray*}
A^1_{2.2} &=& \langle t \partial_t +x \partial_x, t^2 \partial_t +x^2 \partial_x+ mut \partial_u \rangle \ (m \in  R ): \\
&& g = [mt+(k-m)x ]t^{-1} (t-x)^{-1}, \ k\not =0, \\
&& f = |t-x|^{m-2} |x|^{-m} \tilde f (\omega), \\
&& \omega = u |t-x|^{-m} |x|^m , \tilde f_{\omega \omega} \not =0; \\
A^2_{2.2} &=& \langle  t \partial_t +x \partial_x, t^2 \partial_t +mtu \partial_u \rangle \ (m \in  R ): \\
&& g = t^{-2} [kx +mt], \ k \not =0, \ f = |t|^{m-2} |x|^{-m} \tilde f(\omega), \\
&& \omega = |t|^{-m} |x|^m u, \ \tilde f_{\omega \omega} \not =0; \\
A^3_{2.2} &=& \langle t \partial_t +x \partial_x, x^2 \partial_x +tu \partial_u \rangle : \\
&& g = (tx)^{-1} (mx-t) \ (m \in  R ), \ f=x^{-2} e^{-tx^{-1}} \tilde f(\omega), \\
&& \omega = u e^{tx^{-1}}, \ \tilde f_{\omega \omega} \not =0.
\end{eqnarray*}
\end{thm}
In Theorem \ref{t3} we gave all realizations of two-dimensional
Lie algebras which give symmetries of equation (\ref{eq1}): there
are three inequivalent such realizations.

\begin{thm}\label{t22}

There are two equations of the form given in (\ref{L3}) which
admit local one-parameter symmetry groups generated by the
canonical forms given in Theorem \ref{1.32}. They are described by
the following list, where $\langle Q\rangle$ denotes the algebra
generated by the operator $Q$ and we define the equation of the
form (\ref{L3}) by the form of the functions $f(t,x,u)$:
\begin{eqnarray*}
A^1_1 &=& \langle \partial_t +\partial_x +\epsilon u \partial_u \rangle  \ (\epsilon=0,1): \ f = e^{\epsilon t} \tilde f(\theta, \omega), \\
&& \theta = t-x, \ \omega = e^{-\epsilon t} u; \tilde f_{\omega \omega} \not =0;\\
A^2_1 &=& \langle \partial_t +\epsilon u \partial_u \rangle \ (\epsilon =0,1): f = e^{\epsilon t} \tilde f (x, \omega), \\
&& \omega = e^{-\epsilon t} u, \ \tilde f_{\omega \omega} \not =0.
\end{eqnarray*}
\end{thm}
\newpage
The results of the group classification of the equation (\ref{L3})
has been done in following table.
 The forms of the functions\ $f$\
determining the corresponding
invariant equations (\ref{L3}) are given as follows:\\[3mm]

\begin{tabular}{|c|c|c|c|}\hline
Number & Function $f$& Symmetry operator& Algebra
\\ \hline
1& $e^t \tilde f(\omega), $ &$\partial_t+u \partial_u, \partial_x$&$A_{2.1}$ \\
& $\omega = u e^{-t} , \tilde f_{\omega \omega} \not =0$ & & \\
\hline
2& $e^{t+x} \tilde f(\omega), $ &$\partial_t+u \partial_u,$&$A_{2.1}$ \\
& $\omega = u e^{-t-x} , \tilde f_{\omega \omega} \not =0$ &$
\partial_x+u  \partial_u$ & \\ \hline
3& $(t-x)^{-3} \tilde f(\omega), $ &$-t\partial_t-x \partial_x+u \partial_u,$&$A_{2.2}$ \\
& $\omega =(t-x) u, \tilde f_{\omega \omega} \not =0$ &$
\partial_t+  \partial_x$ & \\ \hline
4& $x^{-1} \tilde f(\omega), $ &$-t\partial_t-x \partial_x-u \partial_u,$&$A_{2.2}$ \\
& $\omega =x^{-1} u, \tilde f_{\omega \omega} \not =0$ &$
\partial_t$ & \\ \hline
5& $(t-x)^{-2} \tilde f(u), $ &$\partial_t+ \partial_x,$&$sl (2, \mathbb{R} )$ \\
& $ \tilde f_{u u} \not =0$ &$ t\partial_t+ x \partial_x,$ & \\
& & $t^2 \partial_t +x^2 \partial_x$& \\ \hline
6& $e^{x^{-1}u}$ &$-t\partial_t+x \partial_u,$&$A_{2.2}\oplus A_1$ \\
&  &$ \partial_t,x \partial_x+u \partial_u$ & \\ \hline
7 &$\lambda |x|^{-m-2} |u|^{m+1},$ &$ \partial_t, t \partial_t -\frac{1}{m} u \partial_u,$ &$A_{2.2} \oplus A_1$ \\
& $\lambda \not =0, m\not =0, -,1 -2$ &$ x \partial_x
+\frac{m+1}{m} u \partial_u$ & \\ \hline 8 &$ \tilde f(u), \tilde
f_{u u } \not =0$ &$ \partial_t, \partial_x, -t \partial_t -x
\partial_x $&$ A_{3.6}$ \\ \hline
9 &$\lambda |u|^{n+1}, \lambda \not =0, n\not =0, -1$ &$ t\partial_t-\frac{1}{n} u \partial_u$ &$A_{2.2} \oplus A_{2.2}$ \\
& &$ x \partial_x -\frac{1}{n} u \partial_u$ & \\
& &$  \partial_t, \partial_x$ &  \\ \hline
\end{tabular}
\\ [4mm]

So the problem of group classification of one class of quasilinear
equations of hyperbolic type with two independent variables has
been solved completely.

\end{document}